\documentclass[amssymb,aps,pre,preprint]{revtex4}
\input epsf
\usepackage{graphicx}%
\usepackage{dcolumn}
\usepackage{amsmath}

\begin{document}
\draft
\title{Diffusion in a Granular Fluid - Simulation}
\author{James Lutsko}
\affiliation{Center for Nonlinear Phenomena and Complex Systems,\\
Universit\'{e} Libre de Bruxelles\\
Campus Plaine, CP 231\\
1050 Bruxelles, Belgium}
\author{J. Javier Brey}
\affiliation{Fisica Te\'{o}rica, Universidad de Sevilla\\
Apartado de Correos 1065, E-41080 Sevilla, Spain}
\author{James W. Dufty}
\affiliation{Department of Physics, University of Florida, Gainesville, FL 32611}
\date{\today}

\begin{abstract}
The linear response description for impurity diffusion in a granular fluid
undergoing homogeneous cooling is developed in the preceeding paper. The
formally exact Einstein and Green-Kubo expressions for the self-diffusion
coefficient are evaluated there from an approximation to the velocity
autocorrelation function. These results are compared here to those from
molecular dynamics simulations over a wide range of density and
inelasticity, for the particular case of self-diffusion. It is found that
the approximate theory is in good agreement with simulation data up to
moderate densities and degrees of inelasticity. At higher density, the
effects of inelasticity are stronger, leading to a significant enhancement
of the diffusion coefficient over its value for elastic collisions. Possible
explanations associated with an unstable long wavelength shear mode are
explored, including the effects of strong fluctuations and mode coupling.
\end{abstract}

\pacs{PACS number(s):45.70.-n, 45.70.Mg, 05.70.Ln}

\maketitle

\section{Introduction}

\label{sec1}

Attempts to describe granular media in terms of a more fundamental
underlying statistical mechanics have met with considerable success. As a
prototype for this approach, in the preceding paper \cite{dbyl01} standard
linear response methods from normal fluids have been applied, {\em mutatis
mutandis}, to the case of an impurity particle diffusing in an isolated one
component fluid of smooth inelastic hard spheres ($d=3$) or disks ($d=2$) of
diameter $\sigma $. The collisions are characterized by a coefficient of
normal restitution $\alpha $. The isolated system (or with periodic boundary
conditions), is not in the typical Gibbs state as for elastic collisions,
but rather in a time dependent homogeneous cooling state (HCS). It has been
shown in \cite{dbyl01}, and will be elaborated further here, that this time
dependent HCS can be exactly transformed to a stationary state description.
In the present work, this stationary state description is evaluated by
molecular dynamics (MD) simulation to measure the mean square displacement
of the impurity, its velocity autocorrelation function, and the resulting
diffusion coefficient defined in terms of a formal Einstein or Green-Kubo
relation, respectively. For practical purposes, attention is restricted to
the case of self-diffusion, where the mechanical properties of the impurity
are the same as those of the fluid particles.

The velocity autocorrelation function was approximated in \cite{dbyl01}
using a cumulant expansion and also by means of kinetic theory methods. In
the usual first Sonine polynomial expansion for the latter, this leads to a
simple exponential decay in an appropriate dimensionless time scale defined
below. For the numerical results considered here, one further approximation
is made in evaluating the decay rate, namely two-particle velocity
correlations are neglected. In this case, the results are the same as those
obtained from the Enskog kinetic theory. Earlier studies of self-diffusion %
\cite{BRCyG00} show excellent agreement of this theory with MD results at
very low number density $n$, over a wide range of values for $\alpha $. The
present work extends that study to higher densities. More specifically,
three-dimensional systems with densities in the interval $0.1\leq n^{\ast
}\equiv n\sigma ^{3}\leq 0.75$, for $0.5\leq \alpha\leq 1$, will be
considered.

The HCS is known to be unstable under long wavelength perturbations or
fluctuations \cite{GyZ93}. To avoid this problem, the system size in most
cases considered is chosen to be smaller than the critical wavelength. The
latter is a function of the density and inelasticity, decreasing with
increasing values of each. Consequently, at the highest densities and
smallest $\alpha $ a small system of $108$ particles is required. At
moderate densities and inelasticities, a system size of $512$ particles is
used. The agreement of theory and simulation is found to be quite good for
all $\alpha $ at $n^{\ast }\leq 0.25$ and for all densities at $\alpha \geq
0.9$. At higher densities and smaller $\alpha $, significant discrepancies
occur, with the diffusion coefficient obtained from MD being almost an order
of magnitude greater than the theoretical estimate at $n^{\ast }=0.75$ and $%
\alpha =0.5$. The failure of the Enskog theory at high densities is
well-known for the case of elastic collisions, due to cage effects and
correlated binary collisions. However, in that case the diffusion
coefficient from MD is smaller than the theoretical prediction. Therefore,
it is clear that a quite different density mechanism is effective for
inelastic collisions.

It is reasonable to expect that the underlying instability is responsible in
part for the above discrepancies. Although the instability is, in principle,
avoided by the use of small systems, the latter is prone to large
fluctuations. These are not quantitatively significant at $\alpha =1$ but
grow rapidly with decreasing $\alpha $, as can be seen in the noise level
for the kinetic temperature. This will be illustrated later on in Fig.\ \ref%
{fig1}. A related issue is the role of the instability in affecting the
usual mode coupling corrections to the Enskog theory. Such mode coupling
terms arise from correlations of spontaneous fluctuations in the long
wavelength hydrodynamic modes, which are significantly modified by the
potential instability. Enhanced fluctuations will be addressed in the
qualitative analysis of the data below.

The plan of the paper is as follows. In Sec.\ \ref{s2}, the relevant results
in the preceding paper are shortly summarized, and a new scale
transformation, useful for practical purposes, is introduced. The equations
of motion of the new phase-space variables lead in a direct way to a
steady-state simulation method for the HCS. Starting from an arbitrary
initial condition, the system rapidly approaches a steady state, whose
properties are simply related to those of the HCS. It is important to stress
that this relationship is not approximate, but an {\em exact} consequence of
a change of variables. In particular, the method to measure the mean square
displacement of a tagged particle as well as its velocity autocorrelation
function is discussed in detail. Moreover, both quantities are shown to lead
to equivalent results for the self-diffusion coefficients. Although in
practice some discrepancies appear in the numerical results obtained by the
two procedures, their origin is well understood.

Comparison of theory and simulation is also started in Sec.\ \ref{s2}, and
completed in Sec.\ \ref{s3}. As mentioned above, the agreement is fairly
good at low densities, but relevant effects, which are not taken into
account by the theory developed in \cite{dbyl01}, show up even at moderate
density. The peculiar nature of these effects for inelastic collisions is
discussed. Numerical evidence is provided indicating that the underlying
hydrodynamic instability associated to the shear mode plays an essential
role in the ``anomalous'' behavior of the self-diffusion coefficient.
Nevertheless, such changes in the diffusion constant do not compromise the
existence of the diffusion process, which is confirmed for such conditions
by simulation results for the mean square displacement. Finally, Sec.\ \ref%
{s4} contains a short summary of the results and also some indications of
the possible extensions of the reported work.

\section{Steady-state simulation method}

\label{s2}

To investigate the nature of diffusion in the HCS and to test the
theoretical results presented in the preceding paper \cite{dbyl01}, a more
fundamental description via MD simulation will be considered. The HCS
distribution function for a fluid of $N+1$ particles has the scaling form 
\begin{equation}
\rho _{hcs}(\Gamma ,t)=\left[ \ell v(t)\right] ^{-d(N+1)}\rho _{hcs}^{\ast
}\left( \{{\bf q}_{ij}/\ell ,{\bf v}_{i}/v(t)\}\right) .  \label{2.1}
\end{equation}%
Here and in the following the notation is the same as in ref.\ \cite{dbyl01}%
. In particular, $v(t)$ is the thermal velocity defined in the usual way
(with Boltzmann's constant set equal to one) and $\ell $ is proportional to
the mean free path of the gas. The time dependence of the above distribution
is due to collisional cooling and is determined from 
\begin{equation}
\partial _{t}v(t)=-\frac{1}{2}\zeta (t)v(t),  \label{2.2}
\end{equation}%
where $\zeta (t)$ is the cooling rate. A direct simulation of the cooling
fluid, as described by the Liouville dynamics in the actual phase variables,
is difficult, since the rapid cooling of the fluid leads to numerical
inaccuracies very soon. One method of dealing with this would be to
periodically redefine the time-scale of the simulation, so that the typical
particle velocity remains of the order of unity. However, if the rescaling
is only used to control the numerical stability of the simulation, the state
being simulated would nevertheless be time-dependent. For this reason, some
simulation studies employ different type of thermostats, such as an
externally imposed Brownian force or thermal boundaries, in order to
generate a steady state. While these methods provide more or less realistic
models of various experimental procedures, they obviously probe a state
which is in some way related to, but not identical with, the homogeneous
cooling state.

Soto {\em et al.} \cite{SyM01}, noting the fact that there is no intrinsic
time scale in the dissipative hard sphere model, have proposed to rescale
all particle velocities after every collision, thus establishing a steady
state similar to that described in \cite{dbyl01}. However, this procedure
has the effect of replacing the binary collision dynamics of the dissipative
hard sphere model by an $N$-body dynamics, since the result of a collision
is to alter {\em all} atomic velocities and not just those of the colliding
atoms. It is reasonable to imagine that the binary dynamics is recovered in
the infinite system limit, but the connection between the HCS and this
dynamics is not clear for the case of the small systems considered in most
of the MD simulations. Instead, we follow the procedure used in ref. \cite%
{Lutskopre}, according to which an {\em exact} mapping of the HCS onto a
steady state is exploited as the basis of the simulation method. Following
the ideas developed in \cite{dbyl01}, all velocities are scaled relative to $%
v(t)$ and the new dimensionless time is given by 
\begin{equation}
ds(t)=v(t)dt/\ell .  \label{2.3}
\end{equation}%
This time scale is a measure of the average collision number. The
corresponding Liouville equation in these variables supports a stationary
HCS solution given by $\rho _{hcs}^{\ast }\left( \{{\bf q}_{ij}^{\ast },{\bf %
v}_{i}^{\ast }\}\right) $, where ${\bf q}_{i}^{\ast }={\bf q}_{i}/\ell $ and 
${\bf v}_{i}^{\ast }={\bf v}_{i}/v(t)$. Moreover, the average value of any
phase function $A(\{{\bf q},{\bf v}\})$ is given by Eq.(41) of reference %
\cite{dbyl01}.%
\begin{equation}
\langle A;t\rangle =\int d\Gamma ^{\ast }\rho ^{\ast }(\Gamma ^{\ast
})A(\{\ell {\bf q}_{i}^{\ast }\left( s\right) ,v(t){\bf v}_{i}^{\ast }\left(
s\right) \}).  \label{1}
\end{equation}%
The dynamics in the new phase space is obtained from%
\begin{equation}
\partial _{s}{\bf q}_{i}^{\ast }\left( s\right) ={\bf v}_{i}^{\ast }\left(
s\right) ,\hspace{0.3in}\partial _{s}{\bf v}_{i}^{\ast }\left( s\right) =%
\frac{1}{2}\zeta ^{\ast }{\bf v}_{i}^{\ast }\left( s\right) +L^{\ast }{\bf v}%
_{i}^{\ast }\left( s\right) ,  \label{2}
\end{equation}%
where $L^{\ast }$ is the dimensionless Liouville operator, and $\zeta ^{\ast
}=\ell \zeta (t)/v(t)$ is the dimensionless cooling rate. This is the same
as the usual dynamics for hard spheres, except that the term proportional to 
$\zeta ^{\ast }/2$ represents an acceleration between collisions which
balances the energy lost during collisions, thus enabling a steady state.

Although it is possible to relate average values of the relevant dynamical
functions of the original variables with the average values of the same
functions of the scaled variables for a general situation, attention will be
restricted in the following to the HCS. For homogeneous systems, $\zeta
^{\ast }$ becomes time independent. In principle, the scaled dynamics
defined by Eq.\ (\ref{2}) can be simulated for the properties of interest,
without the complications of continuous cooling in real time. In this
formulation, the simulation is qualitatively similar to that for elastic
collisions, in the sense that the system rapidly approaches a state for
which the time-average of the instantaneous scaled temperature is constant,
and subsequent determination of average properties is simpler and
numerically more accurate. A technical complication is the need to know the
exact value of $\zeta ^{\ast }$ {\em a priori}, which in general is not
possible, since it is determined by the original dynamics. Consequently, it
is useful to make a second change of scale, 
\begin{equation}
{\bf q}_{i}^{\ast \ast }={\bf q}_{i}^{\ast },\quad {\bf v}_{i}^{\ast \ast
}=\left( w^{\ast }/\zeta ^{\ast }\right) {\bf v}_{i}^{\ast },\quad s^{\ast
}=\left( \zeta ^{\ast }/w^{\ast }\right) s,\quad A^{\ast \ast }\left(
s^{\ast }\right) =A\left( s\right) ,  \label{3}
\end{equation}%
where $w^{\ast }$ is an arbitrary time independent dimensionless frequency.
Equation (\ref{2}) then becomes%
\begin{equation}
\partial _{s^{\ast }}{\bf q}_{i}^{\ast \ast }\left( s^{\ast }\right) ={\bf v}%
_{i}^{\ast \ast }\left( s^{\ast }\right) ,\hspace{0.3in}\partial _{s^{\ast }}%
{\bf v}_{i}^{\ast \ast }\left( s^{\ast }\right) =\frac{1}{2}w^{\ast }{\bf v}%
_{i}^{\ast \ast }\left( s^{\ast }\right) +L^{\ast \ast }{\bf v}_{i}^{\ast
}\left( s^{\ast }\right) ,  \label{4}
\end{equation}%
with $L^{\ast \ast }$ the same $L^{\ast }$, but with the replacements $\{%
{\bf q}_{i}^{\ast },{\bf v}_{i}^{\ast }\}\rightarrow \{{\bf q}_{i}^{\ast
\ast },{\bf v}_{i}^{\ast \ast }\}$. This is the form most convenient for MD
simulation, with any reasonable choice for $w^{\ast }$. More explicitly, the
particle dynamics implied by Eq.(\ref{4}) and implemented in our
event-driven simulations consists of an accelerating streaming between
collisions 
\begin{eqnarray}
\partial _{s^{\ast }}{\bf q}_{i}^{\ast \ast } &=&{\bf v}_{i}^{\ast \ast }, \\
\partial _{s^{\ast }}{\bf v}_{i}^{\ast \ast } &=&\frac{1}{2}w^{\ast }{\bf v}%
_{i}^{\ast \ast },  \nonumber
\end{eqnarray}%
while the effect of the collision of two particles is to alter their
relative velocity according to 
\begin{equation}
{\bf g}_{ij}^{\ast \ast }\rightarrow \widetilde{{\bf g}}_{ij}^{\ast \ast }=%
{\bf g}_{ij}^{\ast \ast }-\left( 1+\alpha \right) \left( \widehat{{\bf q}}%
_{ij}^{\ast \ast }\cdot {\bf g}_{ij}^{\ast \ast }\right) \widehat{{\bf q}}%
_{ij}^{\ast \ast },
\end{equation}%
where ${\bf g}_{ij}^{\ast \ast }={\bf v}_{i}^{\ast \ast }-{\bf v}_{j}^{\ast
\ast }$, $\widehat{{\bf q}}_{ij}^{\ast \ast }$ is the unit vector pointing
from atom $i$ to atom $j$ and the center of mass velocity remains the same.

The relationship of $\zeta ^{\ast }$ to $w^{\ast }$ is determined by the
steady state temperature obtained in the simulation. To see this note that
while the instantaneous kinetic energy of the entire system is clearly not
constant, a corresponding ``temperature'' defined as $T^{\ast \ast
}=d^{-1}\left\langle v_{1}^{\ast \ast 2};s^{\ast }\right\rangle ^{\ast \ast
} $\ does approach a constant, as is shown in Appendix \ref{appA}, 
\begin{eqnarray}
T^{\ast \ast }(s^{\ast }) &=&\frac{1}{2}\left( \frac{w^{\ast }}{\zeta ^{\ast
}}\right) ^{2}\left\{ 1+\left[ \frac{w^{\ast }}{\zeta ^{\ast }\sqrt{2T^{\ast
\ast }(0)}}-1\right] e^{-w^{\ast }s^{\ast }/2}\right\} ^{-2}  \nonumber \\
&\rightarrow &\frac{1}{2}\left( \frac{w^{\ast }}{\zeta ^{\ast }}\right) ^{2}.
\label{5}
\end{eqnarray}%
Thus $\zeta ^{\ast }=w^{\ast }/\sqrt{2T^{\ast \ast }(\infty )}$ which
relates quantities measured during the steady state simulation to those
calculated in \cite{dbyl01}. In summary, simulation of Eqs.(\ref{4}) is
expected to yield a steady state after a short transient period.
Subsequently, ensemble averages of properties can be determined as time
averages by making the usual assumption of ergodicity. These properties are
directly related to those of the HCS by a simple scale transformation as
described above.

This steady-state simulation method removes any limitation on the time for
which trajectories may be followed, but there are other limitations due to a
long wavelength hydrodynamic instability\cite{GyZ93} for systems with \
dimension larger than a critical size $L_{c}=2\pi \ell \sqrt{2\eta ^{\ast
}/\zeta ^{\ast }}$ where $\eta ^{\ast }=\eta (t)/nm\nu (t)\ell $ is the
shear viscosity. These instabilities occur when the decay rate of a shear
mode fluctuation is less than the cooling rate of the thermal velocity so
that their size grows relative to the temperature. In the scaled dynamics,
the instabilities show up as ordinary unstable modes that grow exponentially
with $s^{\ast }$, and are therefore easily identified in the simulations %
\cite{Lutskopre}. Since these growing modes represent a spontaneous breaking
of the (assumed) spatial homogeneity of the system, the impurity particle no
longer undergoes simple diffusion and, therefore, its motion is beyond the
scope of this paper. The critical size $L_{c}$ is a function of the density
and coefficient of restitution, being smaller for high density and small
restitution coefficient. In all the cases reported here, the system size is
smaller than that required for the instability to show up. For a system of $%
108$ atoms and $n^{\ast }\leq 0.5$, this provides no limitation on the
accessible densities and coefficients of restitution, otherwise only
sufficiently low (high) values of the density (coefficient of restitution)
can be studied. A different limitation on the simulation is the ``inelastic
collapse'' \cite{McNamaraYoung}, which occurs for small values of $\alpha $
thereby setting a lower limit on the values accessible to simulation. For
the system sizes considered in this paper, the above mentioned limits have
been characterized in more detail elsewhere \cite{Lutskopre}.

\begin{figure}[tbp]
\centering
\includegraphics[width=5in,height=5in,angle=270]{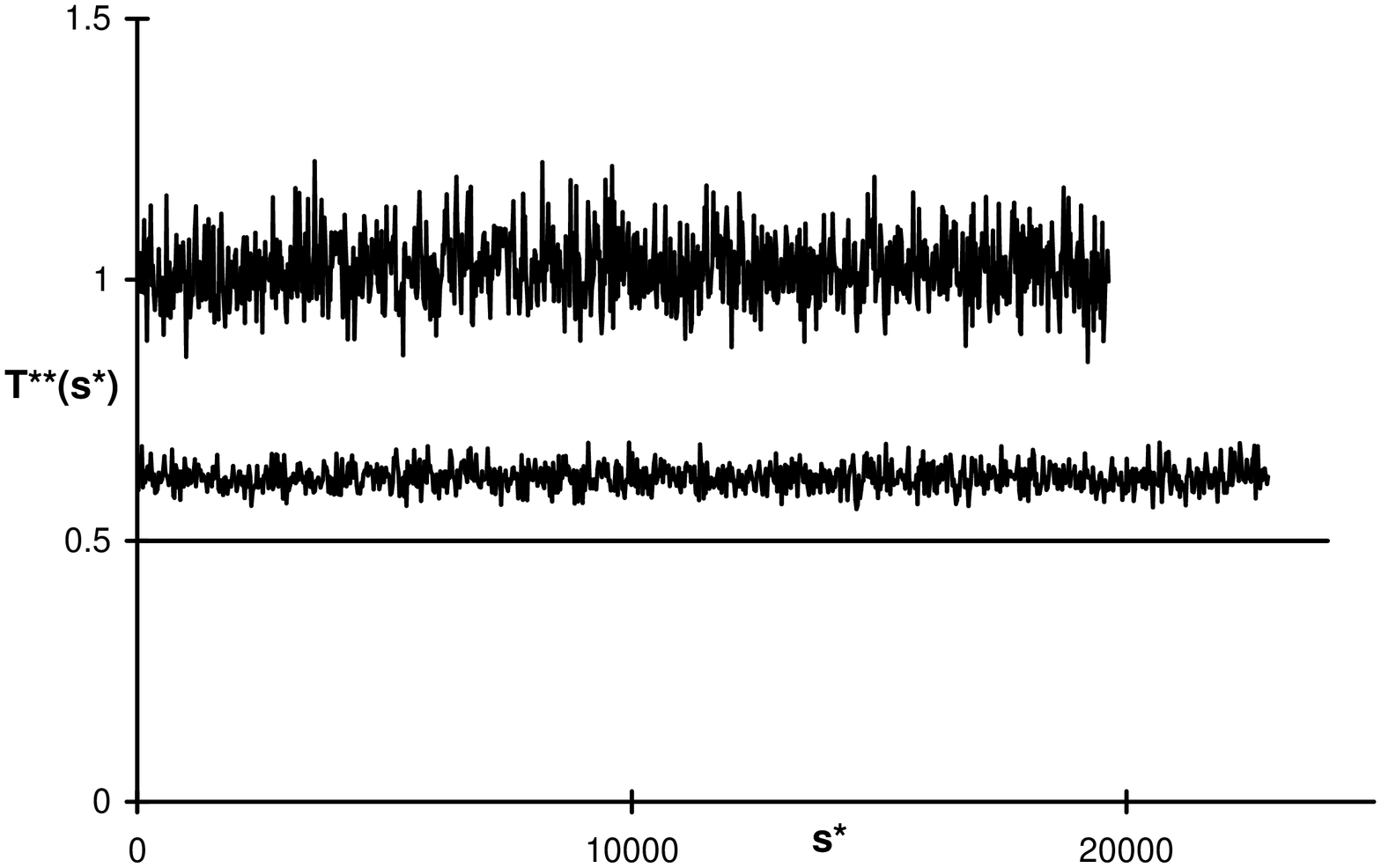}
\caption{Time evolution of the dimensionless steady temperature $T^{\ast
\ast }$ as a function of scaled dimensionless time $s^{\ast}$ for $%
n^{\ast}=0.5$. The lower curve is for $\protect\alpha=1$ , the middle for $%
\protect\alpha =0.7$, and the upper for $\protect\alpha =0.5$.}
\label{fig1}
\end{figure}

The three-dimensional simulations reported in the following begin in all
cases with an equilibrated fluid subject to elastic collisions (i.e. $\alpha
=1,w^{\ast }=0$) with $T^{\ast \ast }=1/2$. The coefficient of restitution
is then set to the desired value and the scaling parameter $w^{\ast }$ is
set to $w^{\ast }=\zeta _{E}^{\ast }\left( \alpha \right) \sqrt{2T^{\ast
\ast }(0)}$ using the Enskog estimate for the cooling rate $\zeta _{E}^{\ast
}\left( \alpha \right) $ as given by Eq.\thinspace\ (74) in ref.\ \cite%
{dbyl01}. This choice ensures that the temperature remains of order $1$ for
all values of $\alpha $, for optimal numerical stability, as well as
provides a continuous path to the equilibrium system for which $\alpha =1$
and $w^{\ast }=0$. The simulation is then continued for $10^{7}$ collisions
to allow the system to reach the steady state at $T^{\ast \ast }=w^{\ast
2}/2\zeta ^{\ast 2}$ where $\zeta ^{\ast }$ is the true cooling rate
generated by the simulation. Since the largest system considered consists of 
$500$ atoms, this corresponds to at least $2\times 10^{4}$ collisions per
atom, and is more than sufficient to reach the steady state. A final
simulation of $10^{7}$ additional collisions is then performed, during which
all statistical averages of interest are accumulated. Figure \ref{fig1}
shows the behavior of the instantaneous kinetic energy (or temperature)
during several typical simulations. In all cases, the instantaneous
temperature fluctuates around a stationary average value, with the size of
the fluctuations increasing as the coefficient of restitution decreases. The
fact that the average value itself varies with $\alpha $ is due to (a) the
inadequacy of the simple estimate of $\zeta ^{\ast }$ we have used for
decreasing values of $\alpha $ and (b) to the presence of long-lived shear
fluctuations, or vortices, which decay on a time scale of thousands of
collisions near the instability. As will be discussed below, the latter give
rise to large fluctuations in all components of the kinetic contribution to
the pressure tensor as well as raising the apparent temperature.

\begin{figure}[tbp]
\centering
\includegraphics[width=5in,height=5in,angle=270]{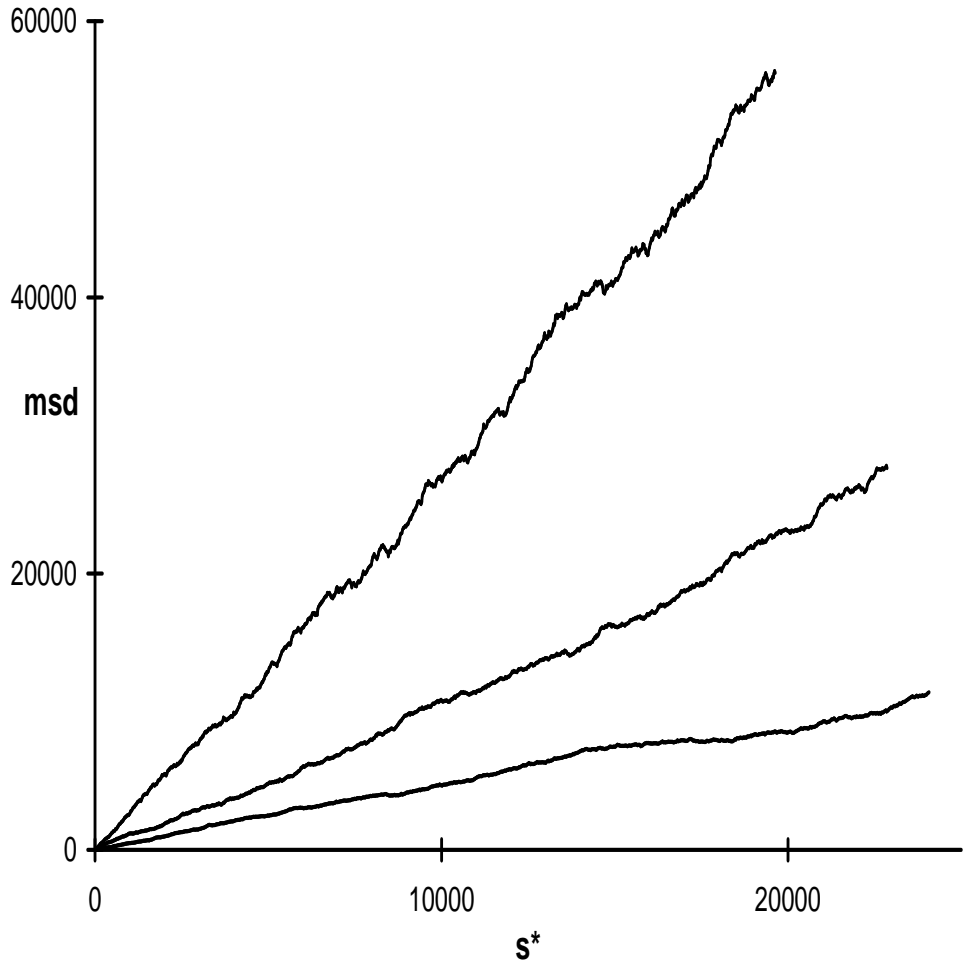}
\caption{Average mean square dimensionless displacement (msd) as a function
of the dimensionless time in the steady-state dynamics for $n^{\ast}=0.5$.
The lower curve is for $\protect\alpha =1$\ , the middle for $\protect\alpha %
=0.7 $, and the upper for $\protect\alpha =0.5$.}
\label{fig2}
\end{figure}

\begin{figure}[tbp]
\centering
\includegraphics[width=5in,height=5in,angle=270]{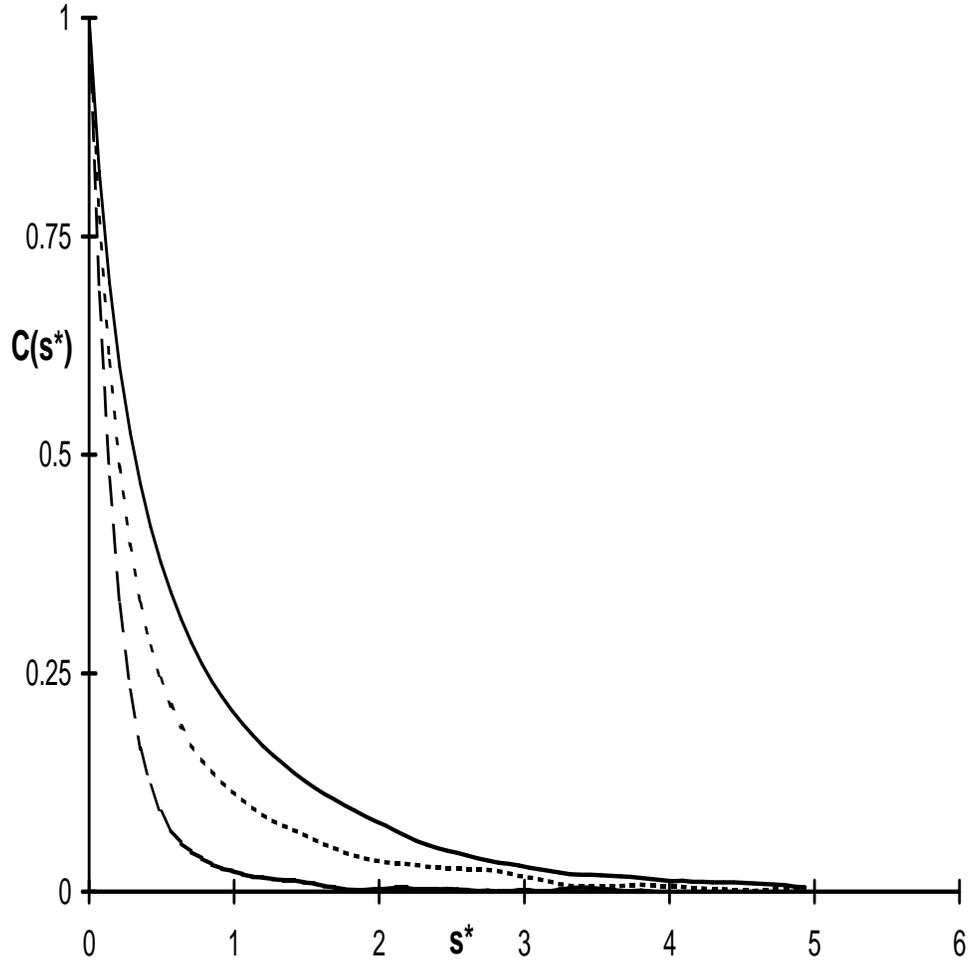}
\caption{Normalized velocity autocorrelation function $C(s^{\ast})$ as a
function of the dimensionless scaled time $s^{\ast}$ in the steady-state
dynamics for $n^{\ast}=0.5$. The values of the coefficient of restitution
are $\protect\alpha =1$\ (lower curve), $\protect\alpha =0.7$ (middle
curve), and $\protect\alpha =0.5$ (upper curve).}
\label{fig3}
\end{figure}

The system is expected to show a typical diffusive behavior under the scaled
dynamics after a short transient time, i.e. for $s^{\ast }>>1$. Then the
diffusion coefficient can be obtained either from the mean squared
displacement (msd) using the Einstein relation, Eq.\thinspace\ (70) of ref. %
\cite{dbyl01}, or from the Green-Kubo relation, Eq.\thinspace\ (69) of ref.%
\cite{dbyl01}. The former is evaluated, for $d=3$, from%
\begin{equation}
D^{\ast \ast }(s^{\ast })=\frac{\zeta ^{\ast }}{w^{\ast }}\,D^{\ast }(s)=%
\frac{1}{6(N+1)}\frac{\partial }{\partial s^{\ast }}\sum_{i=0}^{N+1}\left| 
{\bf q}_{i}^{\ast \ast }(s^{\ast })-{\bf q}_{i}^{\ast \ast }(0)\right| ^{2}.
\label{2.12}
\end{equation}%
Figure \ref{fig2} shows some typical simulation data exhibiting the expected
linearity of the mean-square displacement as a function of $s^{\ast }$. The
Green-Kubo relation expresses the diffusion coefficient in terms of the
velocity autocorrelation function (vacf) as 
\begin{equation}
D^{\ast \ast }(s^{\ast })=\frac{1}{3}\int_{0}^{s^{\ast }}ds^{\ast \prime
}\,\langle {\bf v}_{0}^{\ast \ast }(s^{\ast \prime })\cdot {\bf v}_{0}^{\ast
\ast }\rangle ^{\ast \ast }.  \label{2.13}
\end{equation}%
The vacf is evaluated from%
\begin{equation}
<v_{\alpha }^{\ast \ast }(s^{\ast })v_{\beta }^{\ast \ast }(0)>=\frac{1}{%
(N+1)j_{\max }}\sum_{i=1}^{N+1}\sum_{j=0}^{j_{\max }}v_{i\alpha }^{\ast \ast
}(s^{\ast }+j\Delta )v_{i\beta }^{\ast \ast }(j\Delta ),  \label{2.14}
\end{equation}%
where $\alpha ,\beta ={x,y,z}$, $\Delta $ is the sampling period, and $%
j_{\max }$ is a function of the amount of data available and the time at
which the steady state is established. Notice that the stationarity of the
scaled dynamics is explicitly used to evaluate the vacf for a given time
separation $s^{\ast }$, by averaging over many different samplings during
the simulation. In the simulations, $\Delta $ is taken to be $1/4$ of the
Boltzmann mean free time at $T^{\ast \ast }=1/2$, i.e. $\Delta ^{\ast \ast }=%
\frac{\sqrt{2}}{4}\left( 4n\sigma ^{3}\sqrt{\pi }\right) ^{-1}$. As
discussed above, $T^{\ast \ast }=1/2$ is not the exact value of the
converged temperature, but in formulating $\Delta ^{\ast \ast }$ only a
reasonable order of magnitude is required. To evaluate the right side of
Eq.\ (\ref{2.14}), we follow the procedure of Futrelle and McGinty \cite%
{Futrelle} (see also the discussion in ref.\thinspace\ \cite{Allen}),
according to which the components of the velocity for each atom sampled at
intervals of $\Delta ^{\ast \ast }$ are stored. When the number of values
stored reaches a threshold, $N_{\Delta }$, the convolution theorem is used
to evaluate Eq.\ (\ref{2.14}) by means of the fast Fourier transform, and so
to obtain an estimate of the velocity autocorrelation function. This
procedure is then repeated until the end of the simulation and the various
estimates, typically between tens and hundreds of samples, are averaged to
obtain the final estimate of the correlation function. Figure \ref{fig3}
shows some typical correlation functions obtained in this way, using $%
N_{\Delta }=1024$. The functions have been normalized in all cases by their
initial values, i.e. the function $C_{vv}^{\ast \ast }$, as defined in the
preceding paper, has been plotted. To get the diffusion constant, the
integral of the velocity correlation function is computed using Simpson's
rule and the variance of the pool of estimates is used as the basis for the
calculation of the standard error.

\begin{figure}[tbp]
\centering
\includegraphics[width=5in,height=5in,angle=270]{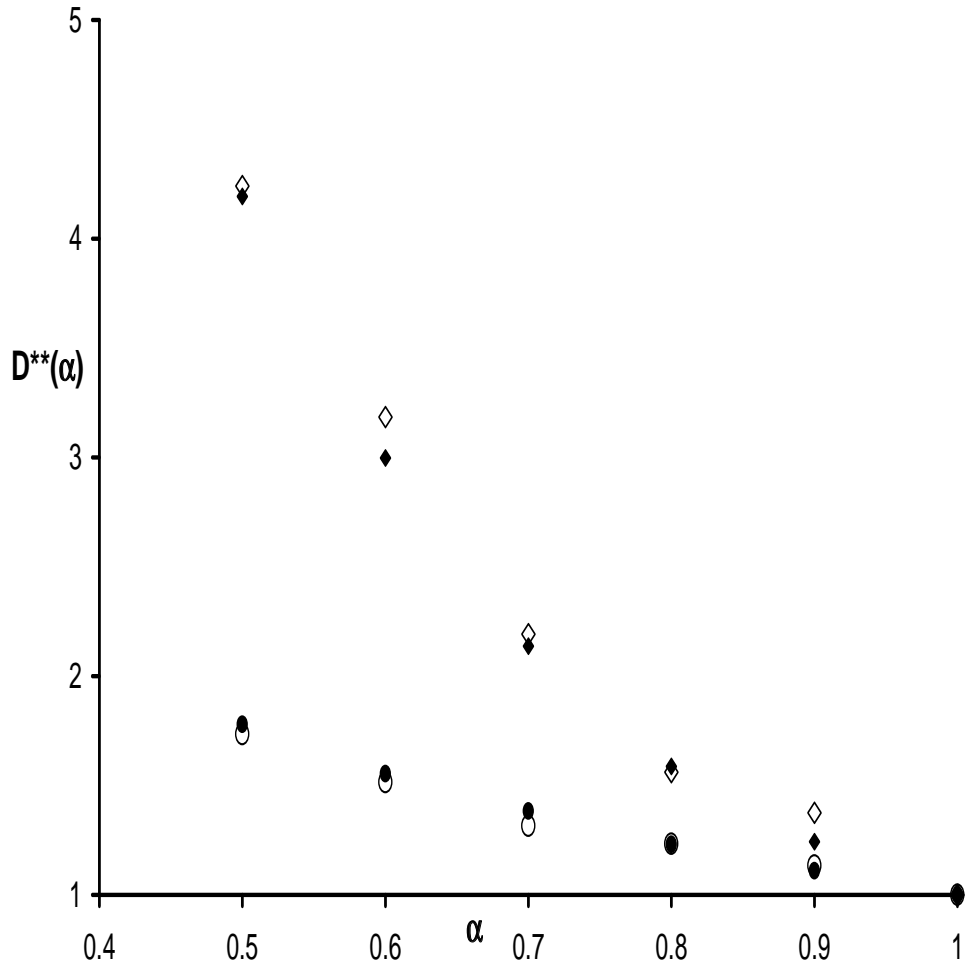}
\caption{Diffusion constant as a function of alpha for $n^{\ast}=0.1$
(circles) and $n^{\ast}=0.5$ (diamonds), as determined from the
mean-squared-displacement (open symbols) and the velocity autocorrelation
function (full symbols). The values are normalized with those obtained for $%
\protect\alpha=1$}
\label{fig4}
\end{figure}

Both methods, Einstein relation and Green-Kubo relation, should yield the
same values of the coefficient of self-diffusion. Figures 2 and 3 show that
the mean square displacement becomes linear in $s^{\ast }$ and the velocity
autocorrelation function decays to zero for $s^{\ast }\succsim 10$. For
larger $s^{\ast }$ both forms for the diffusion coefficient approach a
constant, and all further discussion is restricted to this {\em diffusion
constant}. The consistency between the Green-Kubo and Einstein relations is
confirmed in Fig.\ \ref{fig4}, where the estimates obtained by both
procedures are compared for a system of 108 atoms and several values of the
density and of the coefficient of restitution. While in qualitative
agreement, the difference in values obtained by the two methods is in some
cases as great as $10\%$. It is expected that the results based on the
velocity autocorrelation function are more accurate, because the evaluation
at each fixed time $s^{\ast }$, is based on an average over all times
included in the simulation. In contrast, the mean square displacement is
obtained from a single evaluation for each $s^{\ast }$. This qualitative
difference in accuracy has been confirmed by running multiple simulations
for a few systems, and observing variations in the coefficient of
self-diffusion as large as $10\%$ when determined from the msd, whereas that
determined from the vacf showed variations of less than $1\%$.

\section{Comparison between MD and theory}

\label{s3}

\begin{figure}[tbp]
\centering
\includegraphics[width=5in,height=5in,angle=270]{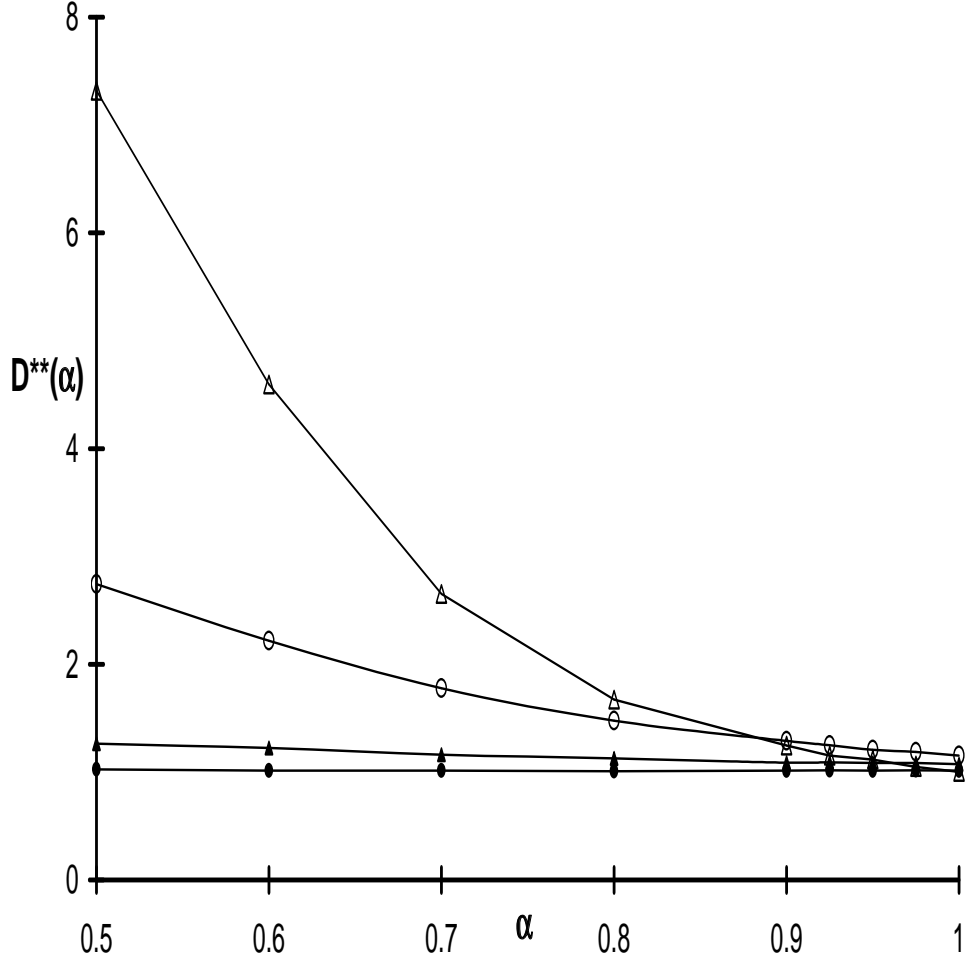}
\caption{Ratio of the diffusion constant determined by MD to the predicted
value based on the Enskog theory, as a function of $\protect\alpha $, for $%
n^{\ast}=0.1$ (full circles), $n^{\ast}=0.25$ (full triangles), $%
n^{\ast}=0.5 $ (open circles), and $n^{\ast}=0.75$ (open triangles). The
lines are a guide to the eye.}
\label{fig5}
\end{figure}

\begin{figure}[tbp]
\centering
\includegraphics[width=5in,height=5in,angle=270]{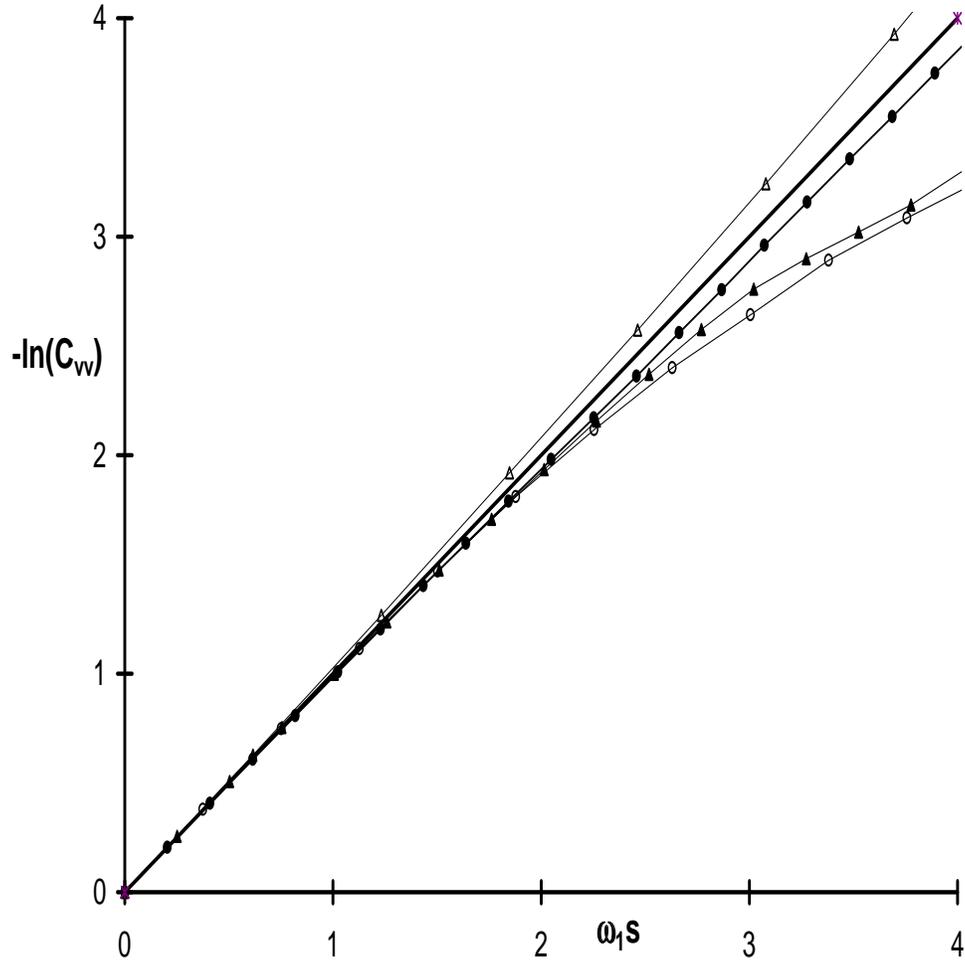}
\caption{Negative logarithm of the normalized vacf $C_{vv}^{\ast \ast}$ as a
function of time for equilibrium for a system in equilibrium ($\protect%
\alpha =1$). Symbols are as in Fig. {\ref{fig5}}. The lines are a guide to
the eye, except the full line, which indicates the Enskog result.}
\label{fig6}
\end{figure}

Figure \ref{fig5} shows the ratio of the observed diffusion constants to the
value predicted by the Enskog-level theory, as a function of $\alpha $ for
densities $n^{\ast }=0.1,0.25,0.5$, and $0.75$. For the lowest density,
theory and simulation are in excellent agreement even at strong dissipation,
as expected from earlier comparisons of simulation results and predictions
of the Boltzmann-Enskog equation \cite{BRCyG00}. As the density increases,
deviations from the Enskog prediction are expected, since they are known to
occur even at equilibrium ($\alpha =1$). In this respect, our results for $%
\alpha =1$ are quite consistent with previous studies of elastic hard spheres%
\cite{ByY80}. It is seen in the figure that deviations increase with
decreasing $\alpha $. For example, at $n^{\ast }=0.25$ the deviations are
less than $10\%$ for $\alpha \geq 0.9$, but increase rapidly to the order of 
$25\%$ for $\alpha =0.5$. This behavior can also be observed in the
normalized vacf itself which, in the Enskog approximation (here and below we
refer to the first Sonine approximation to the Enskog theory), is given by $%
\exp (-\omega _{1}^{\ast }s^{\ast })$, with \cite{dbyl01} 
\begin{equation}
\omega _{1}^{\ast }=\frac{2(1+\alpha )^{2}}{3}\chi \left( \pi T^{\ast \ast
}\right) ^{1/2},  \label{2.15}
\end{equation}%
where $\chi $ is the pair correlation function for two particles at contact.
In Fig.\ \ref{fig6}, the logarithm of the normalized velocity
autocorrelation function, $C_{vv}^{\ast \ast }(s^{\ast })$, is plotted as a
function of $\omega _{1}^{\ast }s^{\ast }$, for the case of {\em elastic}
collisions, $\alpha =1$, and the several densities we have been considering.
Here and below the choice has been made, somewhat arbitrarily but
consistently, of truncating the vacf at $\ln C_{vv}^{\ast \ast }(s^{\ast
})\simeq -4$ in order to eliminate the tails which are dominated by noise.
The data confirm findings of earlier studies \cite{ByY80} that the Enskog
theory gives a good description at low densities, but increasingly
underestimates the diffusion constant, due to the neglect of correlated
collisions and cage effects, at higher densities with a maximum deviation
around $n^{\ast }=0.5$. Above this density, the neglected processes begin to
cancel one another and near $n^{\ast }=0.75$, the diffusion constant crosses
the Enskog prediction, and for higher densities is over-estimated by Enskog
theory. Since the Enskog theory is exact at short times\cite{dbyl01}, these
effects appear as deviations from the simple exponential form of the vacf at
longer times.

\begin{figure}[tbp]
\centering
\includegraphics[width=5in,height=5in,angle=270]{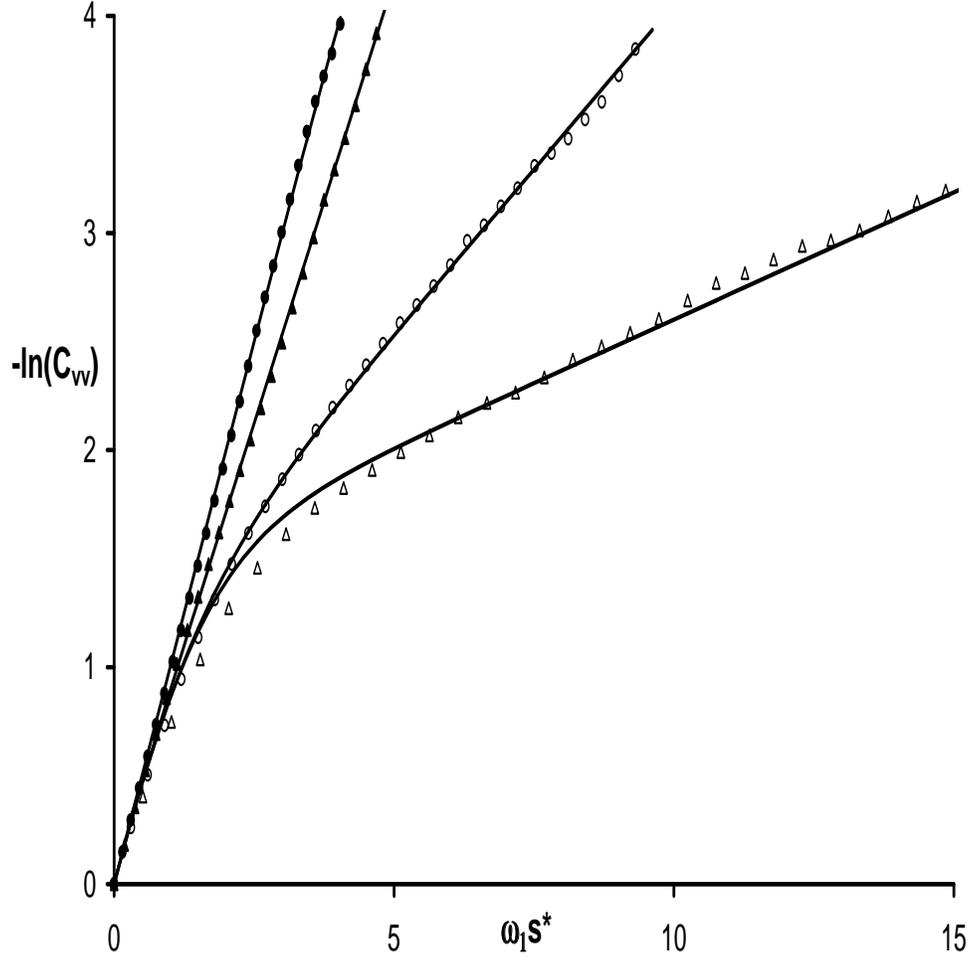}
\caption{Negative logarithm of the normalized vacf $C_{vv}^{\ast \ast}$ as a
function of dimensionless time for the steady-state dynamics with $\protect%
\alpha =0.7$. Symbols are as in Fig.\ {\ref{fig5}}. The lines are the
results of fits of the memory function model with an exponential kernel as
discussed in the main text. The lowest density results are indistinguishable
from the Enskog prediction on this scale.}
\label{fig7}
\end{figure}

Now consider $\alpha <1$. A similar plot, Fig.\ \ref{fig7}, for an inelastic
system with $\alpha =0.7$, again shows good agreement with the Enskog
prediction at the lowest density, but it exhibits much larger deviations
than in the elastic case at higher densities, except at short times where
the Enskog theory is exact, as already pointed out. Interestingly, the data
clearly indicate a crossover to a slower constant decay rate at longer
times. Of course, this behavior increases the time integral of the velocity
autocorrelation function, and it is responsible for the enhancement of the
diffusion coefficient seen in Fig.\ \ref{fig5}, in contrast to the opposite
behavior for high densities at $\alpha =1$.

To characterize the deviation of the vacf from the simple exponential form,
we start with an exact expression for the vacf based on the Zwanzig-Mori
formalism (see Appendix \ref{appB}),%
\begin{equation}
\partial _{s^{\ast }}C_{vv}^{\ast \ast }\left( s^{\ast }\right) +\omega
_{1}^{\ast }C_{vv}^{\ast \ast }\left( s^{\ast }\right) +\int_{0}^{s^{\ast
}}ds^{\prime }\,M(s^{\ast }-s^{\prime })C_{vv}^{\ast \ast }(s^{\prime })=0.
\label{2.16}
\end{equation}%
where $M(s)$ is known as the memory function. If $M(s)$ is neglected the
exponential decay of the above Enskog approximation is recovered, so the
memory function incorporates all of the effects neglected in that
approximation. A simple ansatz for this function as an exponential is
qualitatively successful in modelling the vacf of fluids with elastic
collisions (see \cite{ByY80} and references therein), and can be expected to
work also for inelastic systems, when formulated in the dimensionless time $%
s^{\ast }$. If we substitute $M(s^{\ast })=M(0)\exp (-\lambda s^{\ast })$
into Eq.(\ref{2.16}), and solve for the vacf with the boundary condition $%
C_{vv}^{\ast \ast }(0)=1$, the resulting model is%
\begin{equation}
C^{\ast \ast }(s^{\ast })=\frac{\gamma _{-}-1}{\gamma _{-}-\gamma _{+}}\exp
\left( -\gamma _{+}\omega _{1}s^{\ast }\right) +\frac{\gamma _{+}-1}{\gamma
_{+}-\gamma _{-}}\exp \left( -\gamma _{-}\omega _{1}s^{\ast }\right) ,
\label{2.17}
\end{equation}%
where the constants $\gamma _{+}$ and $\gamma _{-}$ can be related to $M(0)$
and $\lambda $. Figure \ref{fig7} includes the result of fitting these two
parameters to the data, being evident that the above model is able to
capture the crossover from the Enskog behavior at short times to the slower
relaxation for longer times.

\begin{figure}[tbp]
\centering
\includegraphics[width=5in,height=5in,angle=270]{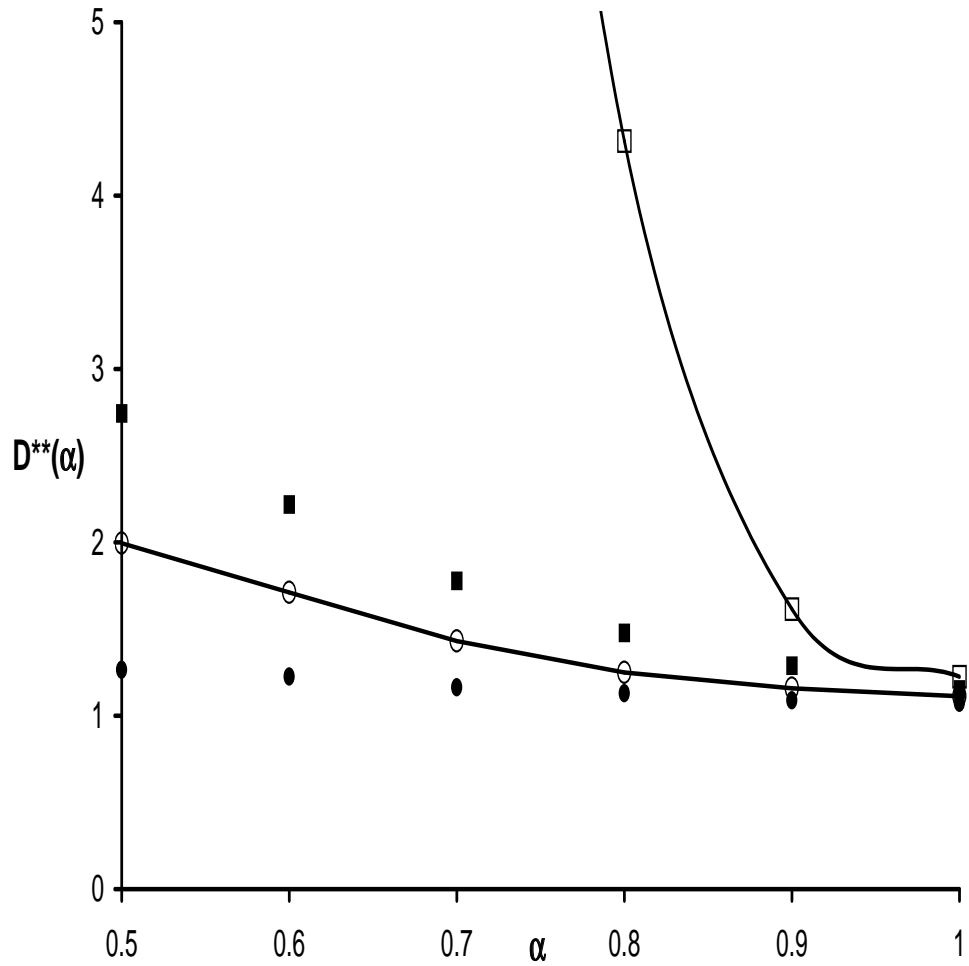}
\caption{Diffusion constants for $n^{\ast}=0.25$ (circles) and $%
n^{\ast}=0.50 $ (squares) as a function of $\protect\alpha $, for 108 atoms
(full symbols) and 500 atoms (open symbols).The lines are a guide to the
eye. }
\label{fig8}
\end{figure}

While the memory function model provides a framework for describing the
results of the simulations, it does not explain them, since any effect not
captured in the Enskog theory will give rise to a non-zero contribution to
the memory function. In a previous study \cite{Lutskopre}, similar
deviations from the Enskog theory were found for the pressure of the system,
and evidence was provided there suggesting that the large discrepancies
between theory and simulation for dense, dissipative states may be due to
the underlying hydrodynamic instability. Additional support for this
possibility comes from Fig.\ \ref{fig8} which shows the diffusion constants
obtained from the simulation data for systems composed by 500 particles
compared to those for systems composed by only 108 particles. For a density
of $n^{\ast }=0.25$, there is a significant further increase of the
diffusion constant in the large system relative to the small system as $%
\alpha $ decreases, with the diffusion constant of the 500 atom system
growing to almost twice that of the 108 atom system at $\alpha =0.5$. At $%
n^{\ast }=0.5$, the enhancement is even larger, but we know \cite{Lutskopre}
that the system is in the unstable regime, at least for $\alpha \leq 0.7$
and possibly for smaller values, so that most of the enhancement is
undoubtedly due to the spontaneous formation of shear flow in the system.
This suggests that in the smaller systems, although they are stable, there
are present large fluctuations characteristic of the instability, such as
spontaneous vortices that form and breakup. In this case, the impurity would
find itself in a fluctuating local flow field which could enhance the
velocity correlations.

\begin{figure}[tbp]
\centering
\includegraphics[width=5in,height=5in,angle=270]{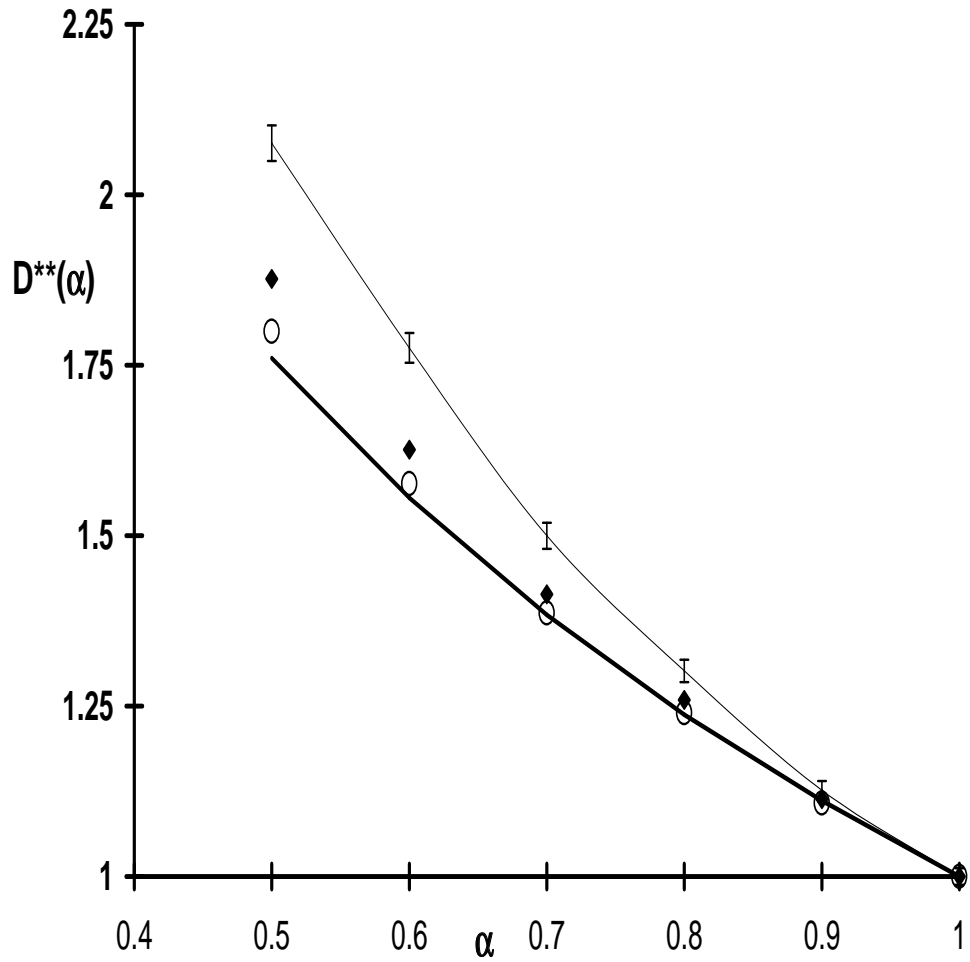}
\caption{Diffusion constants for $n^{\ast}=0.25$ as a function of $\protect%
\alpha $ for 108 atoms, calculated from the Green-Kubo relation (error bars
and connecting line), by removing the longest Fourier modes (diamonds), and
by using the cell-method of computing the local velocity field (circles).The
heavy line is the prediction from Enskog theory.}
\label{fig9}
\end{figure}

\begin{figure}[tbp]
\centering
\includegraphics[width=5in,height=5in,angle=270]{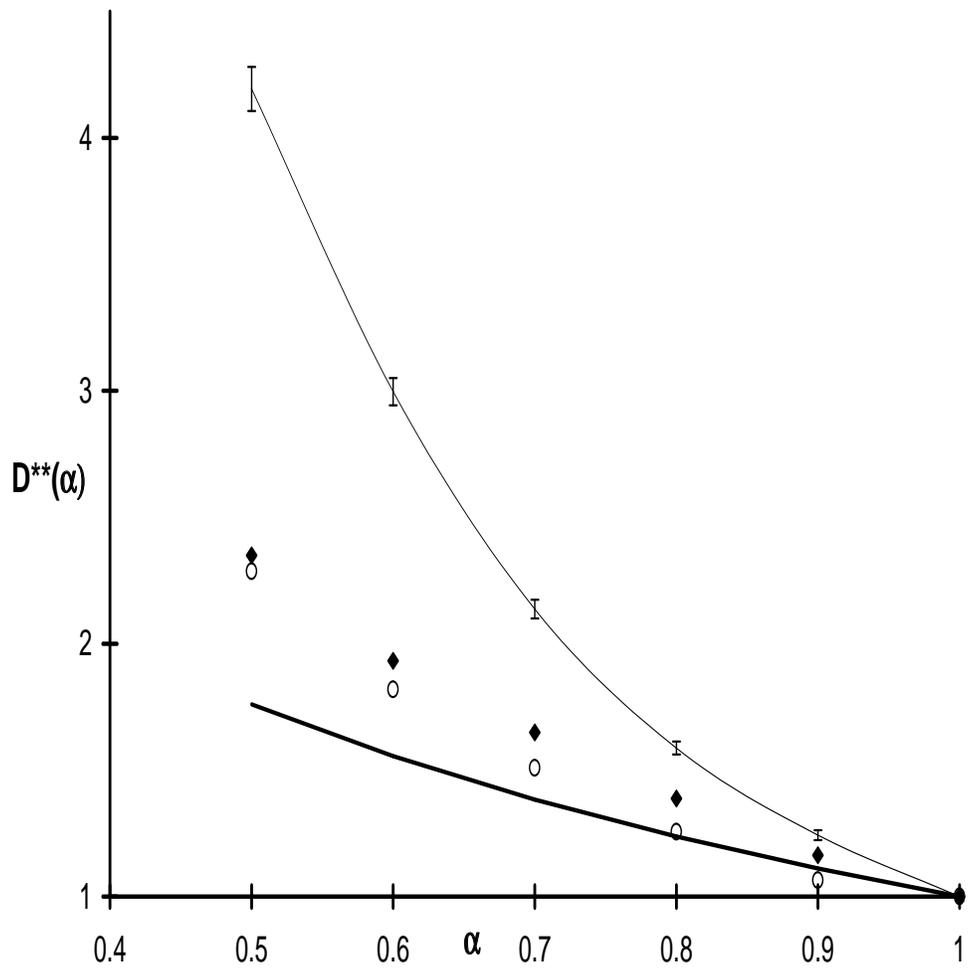}
\caption{The same as Fig.\, {\ref{fig9}} for $n^{\ast}=0.50$.}
\label{fig10}
\end{figure}

\begin{figure}[tbp]
\centering
\includegraphics[width=5in,height=5in,angle=270]{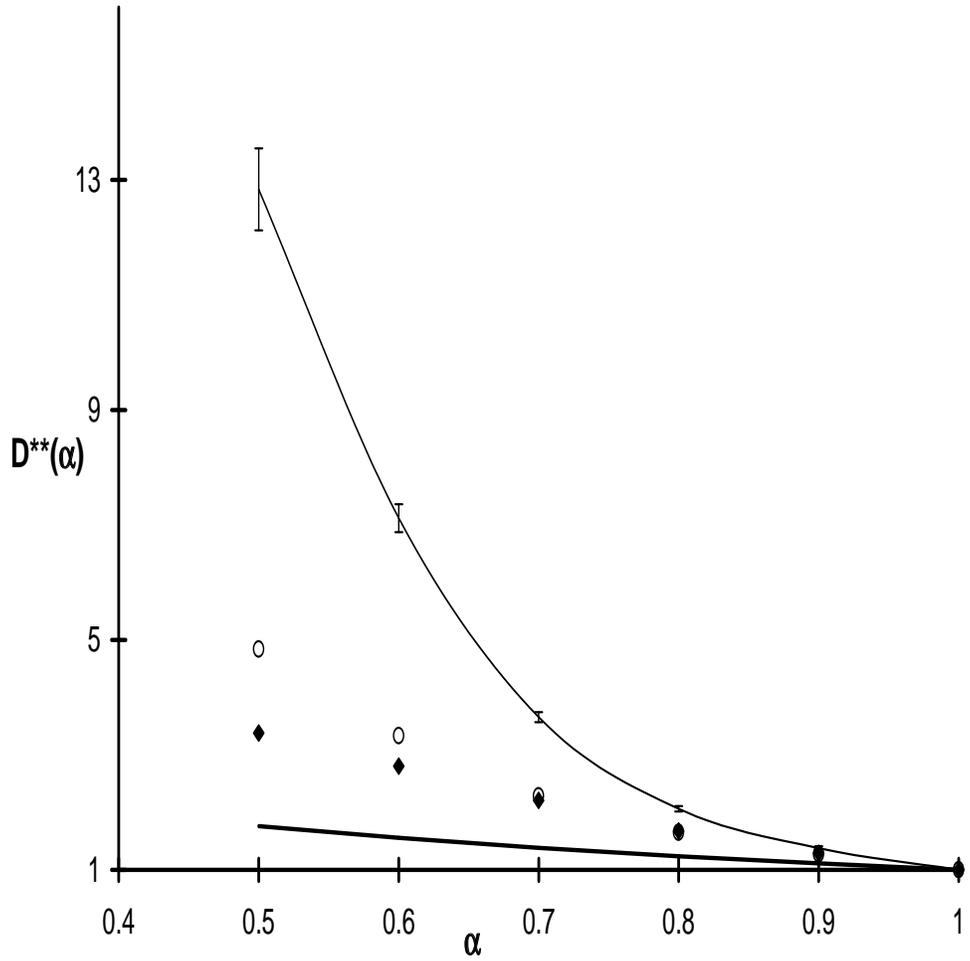}
\caption{The same as Fig.\, {\ref{fig9}} for $n^{\ast}=0.75$.}
\label{fig11}
\end{figure}

To test whether a local flow field plays a relevant role in the
self-diffusion process, two different methods were used to calculate the
autocorrelation of the impurity velocity {\em relative to the instantaneous
local flow field}. In the first method, the instantaneous local flow field
was calculated by dividing the simulation cell into $3^{3}=27$ cubic
subcells. Then each atom's velocity relative to the instantaneous average
velocity of the fluid in the subcell containing it, was used when
calculating both the temperature and the vacf. For the 108 particle system,
this procedure leads to a lowering of the measured value of the equilibrium
diffusion constant by about $10\%$, which we believe to be a finite-size
effect due to the removal of a significant fraction of the degrees of
freedom of the system. For $n^{\ast }=0.5$ it also reduces the steady state
temperature from nearly twice the initial temperature (see Fig. \ref{fig1})
to about $1.5T^{\ast \ast }(0)$ thus demonstrating the size of the
contribution of these fluctuations to the temperature. When the same
procedure is applied to a system of 500 atoms, the shift in the equilibrium
diffusion constant is negligible. The results obtained by this method are
presented in Figs.\ \ref{fig9}-\ref{fig11}. In the graphs, the $\alpha $%
-dependant diffusion constant has been scaled by its measured equilibrium
value. The conclusion emerging from the figures is that the $\alpha $%
-dependence of the new diffusion constant is substantially closer to the
Enskog form. However, this method may be criticized on the grounds that the
statistics of the local velocity field are poor, since the calculations only
involve a few atoms in each cell. We, therefore, also consider a second
method which is specifically designed to eliminate only the part of the
local flow due to the longest wavelength fluctuations in the system. A local
flow field ${\bf u}$ is defined by summing over the smallest Fourier
components compatible with the size and shape of the simulation cell, 
\begin{equation}
u_{\alpha }\left( {\bf r}\right) =\sum_{\left| {\bf k}\right| >2\pi
/L}u_{\alpha }\left( {\bf k}\right) e^{i{\bf k\cdot r}},  \label{2.18}
\end{equation}%
where $L^{3}$ is the volume of the cubic simulation cell. The Fourier
components are determined from the instantaneous velocities of the bulk
fluid via%
\begin{equation}
u_{\alpha }\left( {\bf k}\right) =\frac{2}{N+1}\sum_{i}v_{i\alpha }e^{-i{\bf %
k\cdot r}_{i}}.  \label{2.19}
\end{equation}%
This method is expected to give better statistics than the cell method
discussed above. We then use this flow field to define the relative velocity
used in the calculation of the temperature and the vacf. Figures \ref{fig9}-%
\ref{fig11} also show the resulting diffusion constant when this method of
subtracting the instantaneous fluctuations is used. From the figures follows
that both methods are mutually consistent over the range $0.5<\alpha <1$,
and that the agreement between the simulation results and the theoretical
prediction from Enskog theory is significantly improved. The effects of
subtracting the local flow field are much larger for the inelastic case due
to the mechanism responsible for shear instability.

\section{Discussion}

\label{s4}

In this paper, we have continued the discussion of diffusion in a model
granular system begun in ref.\ \cite{dbyl01}. The main motivation has been
to show that transformation to the steady-state dynamics allows to carry
over many standard methods of nonequilibrium statistical mechanics with
relatively minor modifications. Here, the exact correspondence between the
usual formulation of the dissipative hard-sphere model of granular fluids
and the steady-state dynamics was exploited to formulate a particularly
convenient simulation method which eliminates the need for additional
complications such as exothermic boundary conditions. We have demonstrated
that in the steady-state variables, the homogeneous cooling state exhibits
standard diffusive behavior such as the linear increase of the mean-squared
displacement with time, and the correspondence between the Einstein and
Green-Kubo methods of determining the diffusion constant. In addition, the
simplest approximation of Enskog kinetic theory provides excellent agreement
at low densities for the whole range of inelasticity $0.5\leq \alpha \leq 1$%
. However, although the formalism and concepts of the statistical mechanics
of diffusion have been shown to give an adequate description of diffusion in
the HCS, large quantitative deviations from the Enskog kinetic theory
predictions have been observed even at relatively moderate densities. At the
level of the vacf, it was shown that the deviations are principally due to a
crossover from the Enskog time dependence, which is exact for short times,
to another, slower decay which nevertheless also appears to be exponential
in form. The analytic form of the vacf was shown to be well approximated by
modeling the memory function of the diffusive process by a simple
exponential as has been used in early studies of memory effects in
equilibrium fluids \cite{ByY80}.

Physically, we have shown by means of constrained simulations that the
deviations from the Enskog model are primarily due to the effect of the
longest wavelength velocity modes in the system. For sufficiently large
systems, the shear mode is linearly unstable (in the steady-state picture)
in the classic sense that the time constant associated with its decay goes
to zero as the critical value of $\alpha $ is approached from above. A
linear stability analysis in the steady-state variables\cite{Lutskopre}
shows that the shear fluctuations of wavevector $k$ decay exponentially with
a time constant of $\eta ^{\ast }k^{\ast 2}-\frac{1}{2}\zeta ^{\ast }$,where 
$\eta ^{\ast }$ is the shear viscosity and the stars indicate quantities
expressed in the reduced units of \cite{dbyl01}. The dominant $\alpha $%
-dependence of this expression comes from $\zeta ^{\ast }\sim \left(
1-\alpha ^{2}\right) $ and the size of the system enters through the fact
that the smallest non-zero value of the wavevector the system can sample is $%
k_{\min }^{\ast }=2\pi /L^{\ast }$ where $L^{\ast }$ is the longest
dimension of the simulation cell. For fixed values of $\alpha <1$, there
will always be a critical value of $L^{\ast }$ above which the system is
unstable to shear fluctuations. For values of $\alpha $ above the critical
value, the decay of these fluctuations is nevertheless slowed relative to
equilibrium and we therefore reason that, even before the onset of the
instability and even in systems too small to exhibit the instability, the
slowing down of this mode means that long-lived, long-wavelength
fluctuations are present. These fluctuations give rise to the observed
enhancement of the diffusion constant. This effect is somewhat analogous to
what is observed in turbulent systems since the removal of these modes in
the calculation of the diffusion constant removed a large part of the
discrepancy from the Enskog prediction. The size dependence of the
deviations from the Enskog results supports this conclusion, since larger
systems have a higher critical value of $\alpha $ because the wavevector can
take on smaller values. We conclude that a complete theoretical description
of diffusion in HCS will require a model for the memory function taking into
account the slow relaxation of the shear modes by means, e.g., of a
mode-coupling mechanism.

\bigskip

\section{Acknowledgments}

The research of JWD was supported by National Science Foundation grant PHY
9722133. JJB acknowledges partial support from the Direcci\'{o}n General de
Investigaci\'{o}n Cient\'{\i}fica y T\'{e}cnica (Spain) through Grant No.
PB98-1124. JL acknowledges support from the Universit\'{e} Libre de
Bruxelles.

\appendix

\section{Approach to stationarity}

\label{appA}

The average value of an observable $A(\Gamma )$ for a general homogeneous
state $\rho (\Gamma ,t)$ is given by Eq. (\ref{1}) above which can be
written 
\begin{equation}
\langle A;t\rangle =\int d\Gamma ^{\ast }\rho ^{\ast }(\Gamma ^{\ast
})A(\{\ell {\bf q}_{i}^{\ast }\left( s\right) ,v(t){\bf v}_{i}^{\ast }\left(
s\right) \})
\end{equation}%
and%
\begin{equation}
\partial _{s}{\bf q}_{i}^{\ast }\left( s\right) ={\bf v}_{i}^{\ast }\left(
s\right) ,\hspace{0.3in}\partial _{s}{\bf v}_{i}^{\ast }\left( s\right) =%
\frac{1}{2}\zeta _{hcs}^{\ast }{\bf v}_{i}^{\ast }\left( s\right) +L^{\ast }%
{\bf v}_{i}^{\ast }\left( s\right) .
\end{equation}%
The subscript on $\zeta _{hcs}^{\ast }$ indicates explicitly that it is the
dimensionless cooling rate associated with the HCS%
\begin{equation}
\zeta _{hcs}^{\ast }=\frac{\ell \zeta _{hcs}(t)}{v_{hcs}(t)},
\end{equation}%
where $\ell $ is the mean free path and $v_{hcs}(t)=\sqrt{2T_{hcs}(t)/m}$ is
the HCS thermal velocity. The temperature $T_{hcs}(t)$ obeys the equation 
\begin{equation}
\partial _{t}T_{hcs}(t)=-\zeta _{hcs}(T_{hcs}(t))T_{hcs}(t).
\end{equation}

Now make the change of variables%
\begin{equation}
{\bf q}_{i}^{\ast \ast }={\bf q}_{i}^{\ast },\quad {\bf v}_{i}^{\ast \ast
}=\left( w^{\ast }/\zeta _{hcs}^{\ast }\right) {\bf v}_{i}^{\ast },\quad
s^{\ast }=\left( \zeta _{hcs}^{\ast }/w^{\ast }\right) s,\quad A^{\ast \ast
}\left( s^{\ast }\right) =A\left( s\right) ,
\end{equation}%
to get%
\begin{equation}
\langle A;t\rangle =\int d\Gamma ^{\ast \ast }\rho ^{\ast \ast }(\Gamma
^{\ast \ast },s^{\ast })A(\{\ell {\bf q}_{i}^{\ast \ast },v(t)\left( \zeta
_{hcs}^{\ast }/w^{\ast }\right) {\bf v}_{i}^{\ast \ast }\}),
\end{equation}%
with the definition $d\Gamma ^{\ast \ast }\rho ^{\ast \ast }(\Gamma ^{\ast
\ast },s^{\ast })\equiv d\Gamma ^{\ast }\rho ^{\ast }(\Gamma ^{\ast },s)$.
In particular, the kinetic energy is%
\begin{eqnarray}
\langle \frac{1}{2}mv^{2};t\rangle &=&v_{hcs}^{2}(t)\left( \zeta
_{hcs}^{\ast \ast }/w^{\ast }\right) ^{2}\int d\Gamma ^{\ast }\rho ^{\ast
}(\Gamma ^{\ast \ast },s^{\ast })\frac{1}{2}mv^{\ast \ast 2}  \nonumber \\
&\equiv &\frac{1}{2}mv_{hcs}^{2}(t)\left( \zeta _{hcs}^{\ast \ast }/w^{\ast
}\right) ^{2}\left\langle v^{\ast \ast 2};s^{\ast }\right\rangle ^{\ast \ast
}.
\end{eqnarray}%
The corresponding temperature is%
\begin{equation}
T(t)=\frac{2}{d}\langle \frac{1}{2}mv^{2};t\rangle =2T_{hcs}(t)\left( \frac{%
\zeta _{hcs}^{\ast }}{w^{\ast }}\right) ^{2}T^{\ast \ast }(s^{\ast })
\end{equation}%
where $T^{\ast \ast }(s^{\ast })\equiv d^{-1}\left\langle v^{\ast \ast
2};s^{\ast }\right\rangle ^{\ast \ast }$. The time derivative of $T^{\ast
\ast }(s^{\ast })$ is then found to be%
\begin{equation}
\left( \partial _{s^{\ast }}-w^{\ast }\right) T^{\ast \ast }(s^{\ast
})=-w^{\ast }\frac{\zeta (t)}{\zeta _{hcs}(t)}T^{\ast \ast }(s^{\ast }).
\end{equation}

This is still valid for a general homogeneous state. Now assume the
existence of a scaling solution, which implies $\zeta (t)/\zeta _{hcs}(t)=%
\sqrt{T(t)/T_{hcs}(t)}$. Then 
\begin{equation}
\left( \partial _{s^{\ast }}-w^{\ast }\right) T^{\ast \ast }(s^{\ast })=-%
\sqrt{2}\zeta _{hcs}^{\ast }T^{\ast \ast 3/2}(s^{\ast }).
\end{equation}%
The solution is that given by Eq. (\ref{5}) of the text.

\section{Memory function \ model}

\label{appB}

The dimensionless velocity auto correlation function is defined by

\begin{equation}
C_{vv}^{\ast }(s)\equiv \langle \psi \left( s\right) \psi \rangle ^{\ast },%
\hspace{0.3in}\psi =\frac{v_{0x}^{\ast }}{\sqrt{\langle v_{0x}^{\ast
2}\rangle ^{\ast }}}.
\end{equation}%
Using the notation in ref. \cite{dbyl01} this can be written in terms of the
generators of the dynamics%
\begin{equation}
C_{vv}^{\ast }(s)=\int d\Gamma ^{\ast }\left( e^{{\cal L}^{\ast }s}\psi
\right) \rho _{hcs}^{\ast }\psi =\int d\Gamma ^{\ast }\psi e^{-\overline{%
{\cal L}}^{\ast }s}\left( \rho _{hcs}^{\ast }\psi \right) .
\end{equation}%
The detailed forms for the linear operators ${\cal L}^{\ast }$ and $%
\overline{{\cal L}}^{\ast }$ are given by Eqs. (36) and (42) of ref. \cite%
{dbyl01} but will not be required here. A projection operator is defined by 
\begin{equation}
PX=\rho _{hcs}^{\ast }\psi \int d\Gamma ^{\ast }\psi X.
\end{equation}%
It follows then that 
\begin{equation}
PX(s)=\rho _{hcs}^{\ast }\psi C_{vv}^{\ast }(s),  \label{a4}
\end{equation}%
with the choice 
\begin{equation}
X(s)=e^{-\overline{{\cal L}}^{\ast }s}\left( \rho _{hcs}^{\ast }(\Gamma
^{\ast })\psi \right) .
\end{equation}%
The equation of motion for $X(s)$ is%
\begin{equation}
\left( \partial _{s}+\overline{{\cal L}}^{\ast }\right) X\left( s\right) =0
\end{equation}%
and a closed equation for $PX(s)$ is obtained by operating on this equation
with $P$ and $Q=1-P$ to get the pair of equations%
\begin{eqnarray}
\left( \partial _{s}+P\overline{{\cal L}}^{\ast }P\right) PX\left( s\right)
&=&-P\overline{{\cal L}}^{\ast }QX\left( s\right) ,  \nonumber \\
\left( \partial _{s}+Q\overline{{\cal L}}^{\ast }Q\right) QX\left( s\right)
&=&-Q\overline{{\cal L}}^{\ast }PX\left( s\right) .  \label{a6}
\end{eqnarray}%
Solving formally for $QX\left( s\right) $ in the second equation and
substituting into the first gives the desired closed equation for $PX(s)$%
\cite{Berne}%
\begin{equation}
\left( \partial _{s}+P\overline{{\cal L}}^{\ast }P\right) PX\left( s\right)
-\int_{0}^{s}ds^{\prime }e^{-Q\overline{{\cal L}}^{\ast }Q\left( s-s^{\prime
}\right) }P\overline{{\cal L}}^{\ast }Q\overline{{\cal L}}^{\ast }PX\left(
s^{\prime }\right) =0.  \label{a9}
\end{equation}%
Use of (\ref{a4}) gives the corresponding equation for $C_{vv}^{\ast }(s)$ 
\begin{equation}
\partial _{s}C_{vv}^{\ast }(s)+\omega _{1}C_{vv}^{\ast
}(s)+\int_{0}^{s}ds^{\prime }M\left( s-s^{\prime }\right) C_{vv}^{\ast
}(s^{\prime })=0,  \label{a10}
\end{equation}%
with the definitions 
\begin{eqnarray}
\omega _{1} &=&\int d\Gamma ^{\ast }\psi \overline{{\cal L}}^{\ast }\rho
_{hcs}^{\ast }\psi  \nonumber \\
&=&\langle \left( {\cal L}^{\ast }\psi \right) \psi \rangle ^{\ast }
\label{a11}
\end{eqnarray}%
and%
\begin{equation}
M\left( s\right) =-\int d\Gamma ^{\ast }\psi \overline{{\cal L}}^{\ast }e^{-Q%
\overline{{\cal L}}^{\ast }Qs}Q\overline{{\cal L}}^{\ast }\rho _{hcs}^{\ast
}\psi \text{.}  \label{a13}
\end{equation}%
The definition for $\omega _{1}$ is the same as that for the Enskog
approximation discussed in \cite{dbyl01} and approximated in eq. (\ref{2.15}%
) above. Finally, it is clear that structurally identical expressions would
have been obtained had we performed this derivation using the scaled
variables introduced in Eq. (\ref{3}) of the text thus justifying the
expression in Eq.\ (\ref{2.16}).

\renewcommand{\topfraction}{.99}
\renewcommand{\bottomfraction}{.99}
\renewcommand{\textfraction}{.01}
\renewcommand{\floatpagefraction}{.99}

\clearpage
\clearpage

\clearpage

\clearpage

\clearpage

\clearpage

\clearpage

\clearpage

\clearpage

\clearpage

\clearpage

\clearpage

\end{document}